\documentclass[aps,prl,twocolumn]{revtex4-1}
\usepackage{graphicx}
\usepackage{amsfonts}
\usepackage{amsmath}
\usepackage{color}
\usepackage{bm}
\usepackage{amssymb}
\usepackage{tikz}

\usepackage{multirow}
\usepackage{epstopdf}
\usepackage{natbib}

\newcommand{\beq}{\begin{equation}}
\newcommand{\eeq}{\end{equation}}

\def\R{\mathbf{\hat{R}}}

\def\jcp#1#2#3{{\it J.~Chem.~Phys.}~{\bf #1},\ #2\ (#3)}

\def\prl#1#2#3{{\it Phys.~Rev.~Lett.}~{\bf #1},\ #2\ (#3)}

\def\jpb#1#2#3{J. Phys. B: At. Mol. Opt. Phys. {\bf #1},\ #2\ (#3)}

\def\R{\mathbf{\hat{R}}}

\def\colvecnext#1{
        #1
        \global\advance\colveccount-1
        \ifnum\colveccount>0
                \\
                \expandafter\colvecnext
        \else
                \end{pmatrix}
        \fi
}

\begin{document}

\title{Bayesian optimization for the inverse scattering problem in quantum reaction dynamics}
\author{R. A. Vargas-Hern\'{a}ndez$^1$, Y. Guan$^{2,3}$,  D. H. Zhang$^{2,3}$ and R. V. Krems$^{1,3}$}
\affiliation{\scriptsize 
$^1$Department of Chemistry, University of British Columbia, Vancouver, BC V6T 1Z1, Canada \\
$^2$State Key Laboratory of Molecular Reaction Dynamics, Dalian Institute of Chemical Physics, Chinese Academy of Science, Dalian 116023, China
\\$^3$University of Chinese Academy of Sciences, Beijing 100049, China\\
$^3$Corresponding author: rkrems@chem.ubc.ca
}

\date{\today}

\begin{abstract}
We propose a machine-learning approach based on Bayesian optimization to build global potential energy surfaces (PES) for reactive molecular systems using feedback from quantum scattering calculations. 
The method is designed to correct for the uncertainties of quantum chemistry calculations and yield potentials that reproduce accurately the reaction probabilities in a wide range of energies. 
These surfaces are obtained automatically 
and do not  require manual fitting of the {\it ab initio} energies with analytical functions. 
The PES are built from a small number of {\it ab initio} points by an iterative process that incrementally samples the most relevant parts of the configuration space. 
 Using the dynamical results of previous authors as targets, 
we show that such feedback loops produce accurate global PES with 30 {\it ab initio} energies for the three-dimensional H + H$_2$ $\rightarrow$ H$_2$ + H reaction and 290 {\it ab initio} energies for the six-dimensional OH + H$_2$ $\rightarrow$ H$_2$O + H reaction. These surfaces are obtained from 360 scattering calculations for H$_3$ and 600 scattering calculations for OH$_3$. 
We also introduce a method that quickly converges to an accurate PES without the {\it a priori} knowledge of the dynamical results. By construction, our method illustrates the lowest number of potential energy points (i.e. the minimum information) required for the non-parametric construction of global PES for quantum reactive scattering calculations.   
\end{abstract}

\maketitle

\section{Introduction}

The accurate description of physical processes involving microscopic scattering of molecules is hampered by the lack of knowledge of accurate potential energy surfaces (PES) underlying the scattering events. While sophisticated experiments probing the outcome of molecule - surface scattering  or molecule - molecule collisions can be designed, the theoretical description of such experiments is limited by the difficulty of the electronic structure calculations necessary to compute the microscopic scattering matrices. Instead of computing the PES for molecular scattering from first principles, one can obtain `empirical' PES constructed based on the information about the experimental data and designed to yield an accurate description of the experimental measurements. In the present work, we refer to the construction of such empirical PES as the inverse scattering problem. In general, the inverse scattering problem is a complex task due to the many degrees of freedom relevant for molecular interactions and the complexity of the scattering event. Here, we propose a machine-learning method based on Bayesian optimization for the inverse scattering problem. While the approach is general and can -- in principle -- be applied to any problem with a PES $\leftrightarrow$ observable correlation, we discuss the application of the method to building global PES for chemically reactive molecular systems based on feedback from quantum scattering calculations.

Any quantum dynamics calculation of molecular collision observables involves three steps: (i) computing the potential energy for a wide range of relative atomic coordinates by an {\it ab initio} quantum chemistry method; (ii) fitting these energy points to construct a PES; (iii) integrating the Schr\"{o}dinger equation for the motion of the atomic nuclei on this PES. Unfortunately, it is impossible to compute the potential energy in step (i) without errors and any theoretical predictions of observables are subject to uncertainties stemming from the errors of quantum chemistry calculations.  These uncertainties have a particularly large effect on reactions at low temperatures and make quantitative predictions  
of reaction probabilities at ultracold temperatures impossible \cite{pccp-krems}. 
Therefore, it is necessary to develop approaches that correct the errors of the {\it ab initio} calculations and produce PES yielding the exact match between quantum reaction dynamics calculations and experimental measurements. 


For chemically non-reactive two- or three-atom systems, it was previously shown that empirical PES can be derived from the experimental data \cite{ISP-1,ISP-2,ISP-3,leroy,kirk}. These  approaches generally involve an iterative feedback loop that can be schematically illustrated as follows: 

\begin{minipage}{0.4\columnwidth}
\centering 

\begin{tikzpicture}[->,scale=.7]
   \node (i) at (90:1cm)  {\small (i)};
   \node (j) at (-30:1cm) {\small (ii)};
   \node (k) at (210:1cm) {\small (iii)};
   \draw (70:1cm)  arc (70:-10:1cm);
   \draw (-50:1cm) arc (-50:-130:1cm);
   \begin{scope}[xshift=0.1cm, yshift=0.3cm]
  \draw (210:1cm) arc (180:0:0.8cm);
   \end{scope}
\end{tikzpicture}
\end{minipage} \hspace{4.cm} (1)

\noindent
The PES is computed in step (i), an analytical fit of the PES is generated in step (ii) and the fit is then morphed through a feedback loop involving a series of quantum dynamics calculations. 

Such an approach is  impossible to apply to chemically reactive systems, especially ones involving more than three atoms, for three reasons. First, the active configuration space describing reactive systems is more complex, involving multiple reaction channels, and it is not {\it a priori} known which part of the configuration space is most important for the outcome of a reactive process. Second, each reaction dynamics calculation is time consuming. Third, fitting PES for reactive systems is a complex task that almost always requires manual work \cite{fitting-1,fitting-2,fitting-3,fitting-4,fitting-5}. 

Here, we design a method that makes possible the following optimization loop for three-atom (3D) and four-atom (6D) chemically reactive systems: 

\begin{minipage}{0.4\columnwidth}
\centering
 \begin{tikzpicture}[->,scale=.7]
   \node (i) at (90:1cm)  {\small (i)};
   \node (j) at (-30:1cm) {\small (ii)};
   \node (k) at (210:1cm) {\small (iii)};
   \draw (70:1cm)  arc (70:-10:1cm);
   \draw (-50:1cm) arc (-50:-130:1cm);
   \draw (190:1cm) arc (190:110:1cm);
   
\end{tikzpicture}
\end{minipage} \hspace{4.cm} (2)
\setcounter{equation}{2}

\noindent
This approach allows one to compute and add {\it ab initio} points to the PES incrementally at each iteration. This has four major advantages over conventional approaches to constructing PES:
\begin{itemize}
 \item First, the {\it ab initio} points are only placed in the parts of the configuration space most relevant for the dynamics. The final PES thus offers unique information on the parts of the PES that determine the specific reaction features under study. 
\item Second, this approach eliminates the need for a large number of the {\it ab initio} points, reducing the computational effort associated with quantum chemistry calculations. 
\item Third, as explained below, this approach eliminates the laborious task of fitting the PES. The PES is produced automatically as a mean of a multi-variate distribution. 
\item Finally, the PES thus constructed is guaranteed to yield quantum reaction dynamics observables that agree with the experimental data. 
\end{itemize}

There are two key steps introduced here that make the above optimization loop possible. The reaction observables are approximated by a machine-learning (ML) model, which is a function of another ML model describing the PES. In the present approach, both of these ML models are provided by Gaussian Process (GP) regression \cite{gpbook}. GP regression is a statistical learning technique, which provides a prediction and an uncertainty of the prediction. As described below, the two-tiered ML model is used here to eliminate the need for fitting the PES manually and the uncertainty of the GP prediction is used here to make the optimization loops extremely efficient by means of Bayesian optimization (BO). 

The remainder of the article is organized as follows. We begin by reviewing the application of GP regression to the construction of PES. The following section describes the main ideas behind BO based on GP regression and how  BO can be applied to construct the PES based on the information about the scattering observables. The subsequent sections present the results demonstrating the efficiency of the BO approach.

\section{Gaussian Process Regression for PES}

In our approach, the global reactive PES as a function of the multi-dimensional vector $\bm r$ is given as 
\begin{eqnarray}
V(\bm r) = \sum_{i = 1}^{n} w_i(\bm r) E_i,
\label{sum-representation}
\end{eqnarray}
where $E_i$ are the {\it ab initio} energy points and the weights ($w_i$) are determined by the two-tiered GP model to yield the best outcome of the quantum scattering calculation. 
GPs have been previously used for interpolating PES for molecular dynamics applications  \cite{gp-1,gp-2,gp-3,gp-4,gp-5}, spectroscopic line calculations \cite{jie-jpb, ergei-nn-paper} and molecular scattering calculations  \cite{bo-gao,bowman}. We emphasize that Eq. (\ref{sum-representation}) is not a fit of the PES but a non-parametric regression. 
The coefficients $w_i$ are chosen such that Eq. (\ref{sum-representation}) passes through the potential energy points. Between the points, Eq. (\ref{sum-representation}) represents 
a mean of a distribution that changes in response to the addition of an energy point to the ensemble $\{ E_i \}$ of {\it ab initio} points.  
More specifically, the weights in Eq. (\ref{sum-representation}) are the elements of the following vector \cite{gpbook}:
\begin{eqnarray}
w_i (\bm r_0) = \left [ {\bf A_0^{\intercal}}(\bm r_0){\bf A}^{-1} \right ]_i,
\label{weights}
\end{eqnarray} 
where $\bf A$ is an $n \times n$ matrix of the covariances between all pairs of $E_i$ and $\bf A_0$ is a vector of covariances between the value of energy at $\bm r_0$ and all known $E_i$. The procedure to obtain $\bf A$ and $\bf A_0$ is described elsewhere \cite{jie-jpb,jie-prl} and is briefly summarized in the remainder of this section. 

GP regression is a non-parametric supervised machine-learning algorithm \cite{gpbook}, which belongs to a class of kernel regression algorithms. 
The prediction of a GP is a Gaussian distribution. Essentially, we seek to answer the following question: given the distribution of $n$ {\it ab initio} energy points $E_i$ located at $\bm r_i$ and collectively represented by vector $\bf y$, what is the value of the potential energy at any point $\bm r$ of the configuration space?  GP regression produces a probability distribution for this value. 
This conditional distribution is a Gaussian with the conditional mean $\mu$ and conditional variance $\sigma$ derived \cite{gpbook} to have the following form: 
\begin{eqnarray}
\label{mu}
\mu(\bm r_0) &=& \beta + \bf A_0^\intercal \bf A^{-1} \left ( \bf y - \beta \right )\\
\sigma(\bm r_0) &=& \sigma_u \left (1 - \bf A_0^\intercal \bf A^{-1} \bf A_0 \right )
\label{sigma}
\end{eqnarray}
where $\beta$ and $\sigma_u$ are the unconditional means of the GP. In the present work, we set $\beta$ to zero, assuming that we have no prior knowledge of the PES. This is the simplest and least efficient implementation. Here, we use the mean of the conditional distribution (\ref{mu}) as the value for the interpolation. 

By construction, 
\begin{eqnarray}
\mu(\bm r_i) = E_i
\end{eqnarray}
and 
\begin{eqnarray}
\sigma(\bm r_i) = 0,
\end{eqnarray}
i.e. the Gaussian distributions collapse to a single value at the training points in $\bf y$ so the function $\mu(\bm r)$ passes through the points $E_i$. It is thus an interpolation technique. 

In Eqs. (\ref{mu}) and (\ref{sigma}), $\bf A$ is a square matrix of the covariances between all pairs of energy points $\bm E_i$ in the vector $\bf y$:
\begin{eqnarray}
\mathbf{A} =  {\rm const} \times \left(\begin{array}{cccc} 1 & R(\bm r_1, \bm r_2) & \cdots & R(\bm r_1, \bm r_i) \\  R(\bm r_2, \bm r_1) & 1& \ & \vdots \\ \vdots&  & \ddots &  \\R(\bm r_i, \bm r_1) & \cdots &  & 1 \\\end{array} \right)
\end{eqnarray}
and $\bf A_0$ is  a vector of covariances between the value of the potential energy at $\bm r_0$ and all the values $E_i$ in the vector $\bf y$: 
\begin{eqnarray}
\mathbf{A}_0 =  \left(\begin{array}{c} R(\bm r_1, \bm r_0) \\
R(\bm r_2, \bm r_0) \\
 \cdots \\ R(\bm r_i, \bm r_0)  \\\end{array} \right)
\end{eqnarray}
In order to predict the distribution of possible values of energy at $\bm r_0$, the method thus relies on the entire vector of known values of energy in the entire configuration space. 
Clearly, the effect of the energy points farther away from $\bm r_0$ is smaller than the effect of the energy points nearby. This is accounted for by the mathematical form of the correlation functions $R(\bm r_i, \bm r_j)$, which must decay as the distance between $\bm r_i$ and $\bm r_j$ increases.

The GP model is trained by determining the best covariance matrix $\bf A$ for a given set of values in $\bf y$. To do this, the functions $R(\bm r_i, \bm r_j)$ are approximated by a simple analytical function. For this work we used the \emph{Mat\'ern} function,
\begin{eqnarray}
R(\bm r_i,\bm{r_j}) &=&  \left(1 + \sqrt{5}r(\bm{r_i},\bm{r_j}) + \frac{5}{3}
r^2(\bm{r_i},\bm{r_j}) \right)
\nonumber
\\
\times \exp \left(-\sqrt{5}r^2(\bm{r_i},\bm{r_j})\right)
\label{eqn:matern_k}
\end{eqnarray}
with
\begin{eqnarray}
r^2(\bm r_i ,\bm r_j ) = (\bm r_i -\bm r_j )^{\top} \bm M(\bm r_i -\bm r_j ) \times \nonumber\\
 \begin{pmatrix}
r^1_i - r^1_j, & \cdots, & r^d_i - r^d_j  
\end{pmatrix}
\begin{pmatrix}
\theta_1 &  & \\ 
 & \ddots & \\ 
 &  & \theta_d
\end{pmatrix}\begin{pmatrix}
r^1_i - r^1_j\\ 
\vdots\\ 
r^d_i - r^d_j 
\end{pmatrix}\nonumber
\end{eqnarray}
where $d$ is the number of dimensions and the diagonal matrix $\bm M$ contains the kernel parameters $\theta_i$.  To find the best {\it estimates} of the kernel parameters we maximize the log \emph{marginal likelihood} with respect to the parameters. The log {marginal likelihood} is,
\begin{eqnarray}
\log p({\bf y}|\bm r,\bm{\theta}) = -\frac{1}{2} {\bf y}^\top {\bf A}^{-1} {\bf y} - \frac{1}{2}\log |{\bf A}| -\frac{\it n}{2} \log (2\pi)
\nonumber
\\
\end{eqnarray}
where $|\bf A|$ is the determinant of the matrix $\bf A$ and $n$ is the number of training points.

Note that this procedure does not produce analytical fits of the values in $\bf y$. Neither should the best covariance function $R$ found by maximizing the log-likelihood function be considered an analytical fit of the covariances. Rather, $R$ is an estimator of the covariances. The function $R$ found by training the GP model provides a  function, which  parametrizes the GP predictive distributions. Although the efficiency and the properties (such as the differentiability) of the resulting GP model depend on the choice of the mathematical function for $R$, this choice is not unique and should lead to the same results for large $n$. A simple Gaussian function or an exponentially decaying function could have been chosen in place of the \emph{Mat\'ern} function, leading to the same result, although with a larger $n$ (for the present applications). 

\section{Bayesian Optimization of Scattering Observables}

In the previous section, we described how GP regression can be used for step (ii) in Eq. (2). 
Using GP models for step (ii) provides a way to automatically construct the PES that can be used for quantum dynamics calculations in step (iii). 
Because GP regression produces the PES in the form of Eq. (\ref{sum-representation}), 
this automates the optimization cycle (2), i.e. a different global PES is produced by GP regression for each loop of Eq. (2) without manual work. 
However, each  feedback loop in Eq. (2) involves a scattering calculation, which is often time-consuming. A typical scattering calculation of the reaction observables takes 
minutes to hours of CPU time, depending on the complexity of the reaction system. Therefore, in general, the feedback loop (2) is considered unfeasible. 
In this work we show that such feedback loops can be made very efficient (converging to an accurate surface very quickly) and hence feasible by means of  Bayesian optimization. 


In order to apply Bayesian optimization, we need to construct a machine learning model of the {\it quantum dynamics results}. We do this also by GP regression. The GP model of quantum dynamics results is trained exactly as described in the previous section, with the training points being the results of quantum dynamics calculations instead of the {\it ab initio} energy points.  
This produces a GP model  ${\cal F}[{\cal G}]$ of the scattering calculation results, which is a function of another GP model $\cal G$ giving the PES. 
Once ${\cal F}[{\cal G}]$ is trained, one can apply Bayesian optimization to find the optimal surface $\cal G$. 


The goal of Bayesian optimization (BO) is to find the global minimizer (or maximizer) of an unknown objective function $f$,
\begin{eqnarray}
\bm{x}_* = \text{arg max } [ f(\bm{x}) ] =  \text{arg min} [ - f(\bm{x})].
\end{eqnarray}
BO is a sequential model-based approach originally designed for unknown objective functions for which:  (i) the gradient of $f(\bm{x})$ is not available or difficult to evaluate and (ii) the cost of evaluating $f$ at $\bm{x}$ is high \cite{bo1,bo2}. It is ideally suited for the present application as, in our work, the function $f$ represents the results of quantum scattering calculations, which are generally difficult to compute.

Two ingredients are needed in order to build a BO algorithm. First, we need an emulator function $g(\bm{x})$ that mimics the unknown objective function $f(\bm{x})$.  
In our work, this emulator function will be provided by GP model described in the previous section, but now applied to emulate the quantum dynamics results.  
Second, it is necessary to define an acquisition function $\alpha(\bm{x})$ that determines the policy evaluation of $f(\bm{x})$. In other words, this function will direct the algorithm to the part of the parameter space where the original function must be evaluated. 

The general recipe for the BO algorithm starts with finding the minimum/maximum of the acquisition function ($\bm{x}_* = \text{arg max } \alpha(\bm{x})$). The function $f$ is then evaluated at this position $f(\bm{\bm{x}_*})$ producing an update to the emulator $g$. This procedure is repeated iteratively until the minimizer/maximizer of $g(\bm{x})$ is converged \cite{bo1,bo2}. 

In the present work, we use the following acquisition function 
\begin{eqnarray}
\alpha(\bm{x}) = \mu(\bm{x}) + \kappa\sigma(\bm{x})
\label{aqq}
\end{eqnarray}
where $\mu(\bm{x})$ and $\sigma(\bm{x})$ are the mean and the standard deviation of the Gaussian process $\cal F$. The hyperparameter $\kappa$ balances the trade-off between exploration and exploitation. When $\kappa \gg 1$, the acquisition function stimulates exploration of the global multi-dimensional space because $\alpha(\bm{x}) \approx \sigma(\bm{x})$ so the function directs the evaluation of the original function at parts of the space where the Gaussian process is the least certain. 
On the other hand, when $\kappa \ll 1$ and/or when the uncertainty of the Gaussian process $\sigma$ becomes very small,
the acquisition function follows more closely the mean of the GP model, which results in a more accurate determination of the extremum. 



In this work, we want the algorithm to converge quickly to a substantial (not necessarily global) maximum. Therefore, the value of $\kappa$ must be chosen to be $\ll 1$. The actual value of $\kappa$ determines the efficiency of the algorithm. We choose the value of $\kappa$ by examining the first optimization loop, to ensure that the BO algorithm converges fast and the improved results can be obtained with less than  $15 \times \cal N$ calculations for each iteration, where $\cal N$ is the dimensionality of the configuration space. The effect of the value of $\kappa$ on the convergence speed of BO is discussed in a following section.

\section{Application to quantum reaction dynamics}

We consider two chemical reactions: 
\begin{eqnarray}
{\rm H} + {\rm H}_2 \rightarrow {\rm H}_2 + {\rm H} \\
{\rm OH} + {\rm H}_2 \rightarrow {\rm H}_2{\rm O} + {\rm H}
\end{eqnarray}
We compute the total reaction probabilities for the reactant molecules in the ground ro-vibrational state and the total angular momentum $J = 0$ using the time-dependent wave packet (TDWP) dynamics approach described in Ref. \cite{zhang-1,zhang-2,zhang-3,zhang-4}, explicitly accounting for all  degrees of freedom.  The basis sets of the reaction dynamics calculations are chosen to ensure full convergence. 

Both of these chemical reactions have been studied before \cite{h3, OH3-pes}. For reaction (5), Su {\it et al} \cite{h3} computed the reaction probabilities with the 3D PES from Ref.  \cite{H3-pes}, constructed using an analytical fit to 8701 {\it ab initio} energy points.
 For reaction (6), Chen {\it et al} \cite{OH3-pes} computed the reaction probabilities with the 6D PES constructed using NN fits to $\sim$17 000 {\it ab initio} calculations.   

In conventional approaches, the global PES is constructed before the dynamical calculations. The present approach is conceptually different. 
We begin by randomly selecting a small number of points ($n = 20$ for the 3D surface and $n=280$ for the 6D surface) from the original PESs \cite{H3-pes, OH3-pes} and construct an approximate PES by GP regression, giving Eq. (\ref{sum-representation}) as described in Ref. \cite{jie-jpb}. We denote this GP model of the surface by ${\cal G}(n)$. This model is likely to be highly inaccurate. 



Given ${\cal G}(n)$, we ask: if one {\it ab initio} point is added to the original sample of few points, where in the configuration space should it be added to result in the maximum improvement of the {\it quantum dynamics results}? This formulation corresponds to a class of reinforcement learning strategies in ML \cite{RL,settles}.
In principle, this question could be answered without ML by a series of quantum dynamics calculations based on ${\cal G}(n+1)$ with the added point moved on a grid in the configuration space. However, in practice, such an approach would be completely unfeasible as it would require about $10^{\cal N}$ dynamical calculations for each added {\it ab initio} point, where $\cal N$ is the dimensionality of the configuration space. 

To overcome this problem, we introduce another GP, hereafter denoted by ${\cal F}$,  representing the reaction probabilities as functions of the location of the added {\it ab initio} point.  This GP model is trained by $15 \times {\cal N}$ quantum dynamics calculations. We thus obtain the explicit function of the reaction probabilities ${\cal F} [{\cal G}(n+1, {\bm r})]$, where $\bm r$ is  the $\cal N$-dimensional vector denoting the position of the added {\it ab initio} point in the configuration space. By training $\cal F$,  we also obtain $\Delta {\cal F} [{\cal G}(n+1, {\bm r})]$, representing the uncertainty of $\cal F$ at $\bm r$, as the conditional standard deviation of the GP \cite{jie-jpb}. The GP $\cal F$ and the uncertainty $\Delta {\cal F} [{\cal G}(n+1, {\bm r})]$ allow us to implement BO as described in the previous section. 

We next propose two approaches:  

{\it (1) Fitting known quantum dynamics results.} If the reaction probabilities are known, either from a calculation with the full surface or from an experiment, one can minimize the difference between ${\cal F} [{\cal G}(n+1, {\bm r})]$ and the known results, to obtain the optimal value of $\bm r$. This is the position in the configuration space, where the new {\it ab inito} point must be added. 
This minimization can then be iterated by setting $n + 1 \Rightarrow n$, until $n$ becomes large enough to produce an accurate surface ${\cal G}$. 

{\it (2) Obtaining the best surface without known dynamical results.} If accurate quantum dynamical results are not available, we propose to {\it maximize} the difference of the reaction probabilities computed in two successive iterations: 
\begin{eqnarray}
\sqrt{\{{\cal F} [{\cal G}(n+1, {\bm r})] - {\cal F} [{\cal G}(n)]\}^2},
\label{max-difference}
\end{eqnarray}
while restricting $\bm r$ to be within a physically reasonable range. This is justified by the observation that ${\cal G}( n \rightarrow \infty)$ must produce the best surface so the maximum improvement of the surface at each iteration is achieved when $\bm r$ corresponds to the maximum of Eq. (\ref{max-difference}).

In the subsequent sections, we illustrate these two approaches.

\begin{figure}[!ht]
\includegraphics[width=\columnwidth]{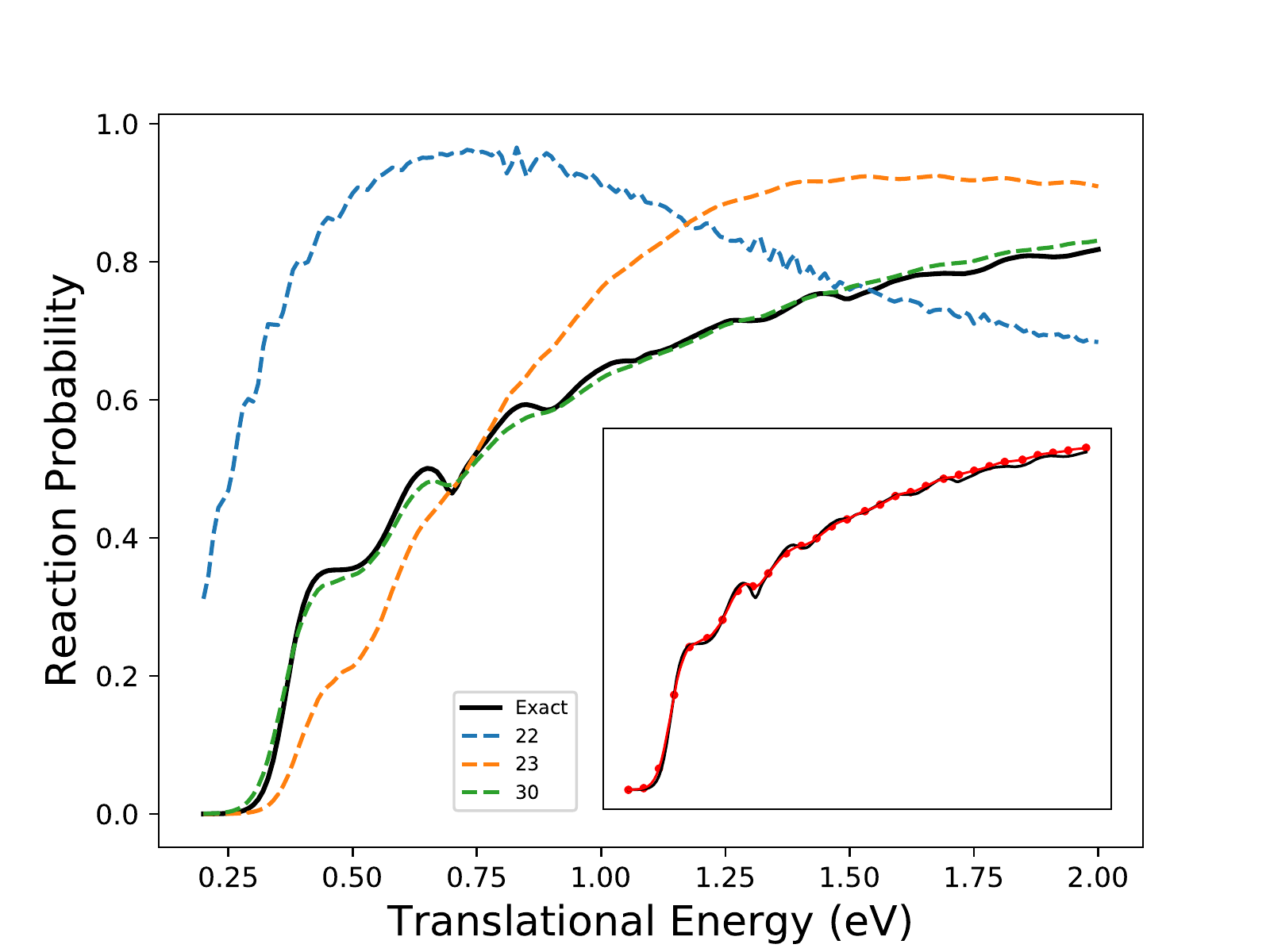} \\
\includegraphics[width=\columnwidth]{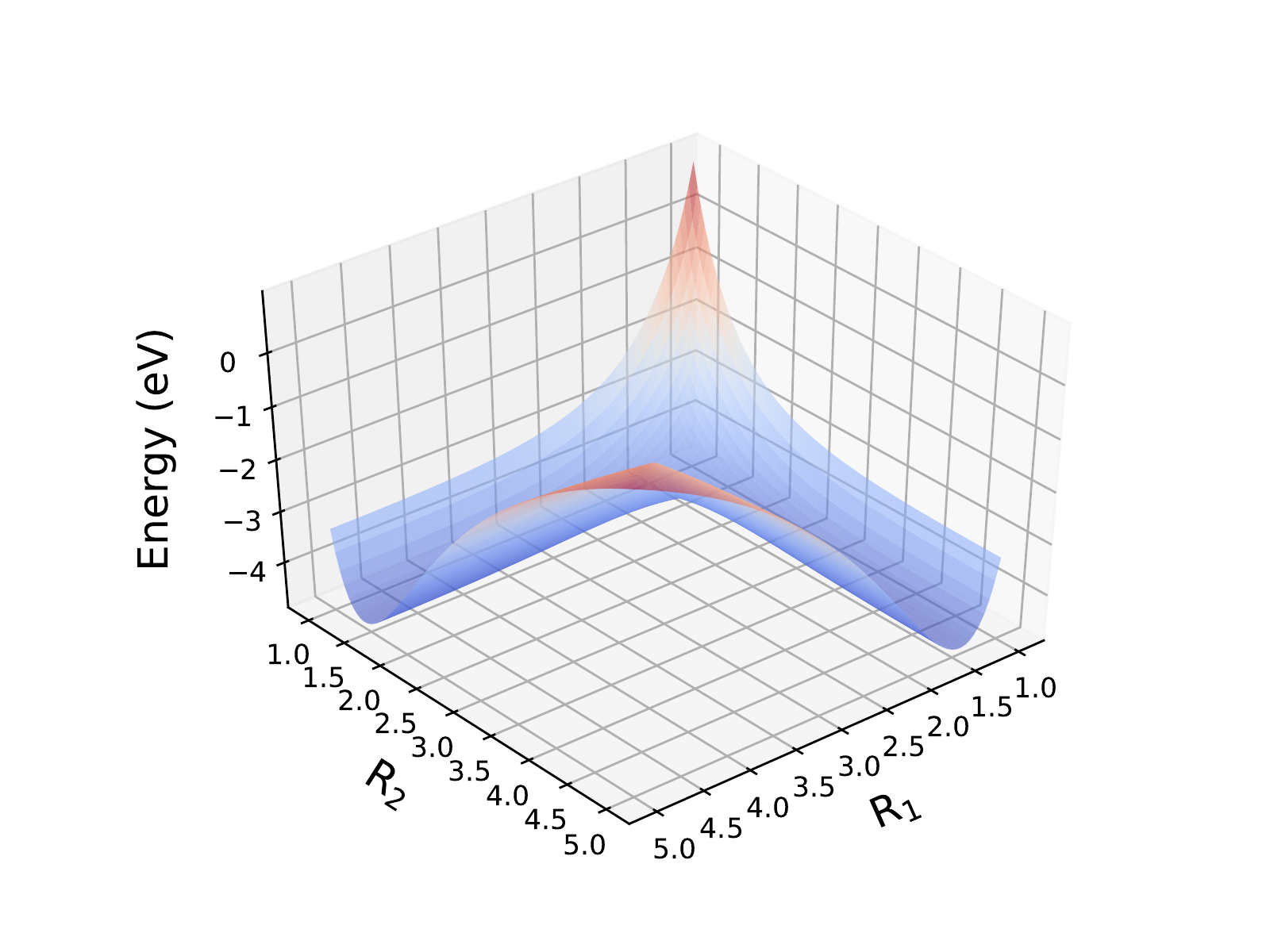}  
\caption{\label{fig1} Upper panel: the reaction probability for the H$_2$ + H $\rightarrow$ H + H$_2$ reaction as a function of the collision energy. The black solid curve -- calculations from Ref. \cite{h3} based on the surface with 8701 {\it ab initio} points from Ref. \cite{H3-pes}. The dashed curves -- calculations based on the GP PES obtained with 22 {\it ab initio} points (blue); 23 points (orange), 30 points (green) and 37 points (inset). The RMSE and maximum error of the results with 37 points are 0.009 and 0.028, respectively. 
Lower panel: The GP model of the PES for the H$_3$ reaction system constructed with 30 {\it ab initio} points.  
$R_1$ and $R_2$ are the distances between atoms 1 and 2 and atoms 2 and 3, respectively.  
}
\end{figure}

\section{Bayesian optimization efficiency}

We begin by illustrating the efficiency of BO for constructing the PES from feedback loops in Eq. (2). Because every cycle in Eq. (2) involves a quantum scattering calculation, the particular parameter of interest is the number of feedback loops necessary to produce a PES yielding accurate quantum dynamical results. 

In this section, our goal is to obtain known reactive scattering probability for reactions (15) and (16), starting from a random guess of a PES and using as few cycles of Eq. (2) as possible. 
To do this, we train ${\cal F}[{\cal G}(n+1, {\bm r})]$ to represent the root-mean-square (RMS) deviation of the reaction probabilities from the known results.  Here, BO amounts to minimizing the following function: 
\begin{eqnarray}
{\cal F} + \kappa \Delta {\cal F}.
\label{F-and-DeltaF}
\end{eqnarray} 
The value of $\kappa$ determines the efficiency of the optimization convergence. We set $\kappa = 0.005$.
As described above, this value of $\kappa$ is chosen by examining the first iteration of cycle (2) to ensure that the algorithm produces improved results with less than $15 \times {\cal N}$ dynamical calculations. 
After a local minimum of $\cal F$ is reached,  the second term in Eq. (\ref{F-and-DeltaF}) directs the minimization algorithm towards the parts of the configuration space where $\cal F$ is least accurate, thus sampling the entire space in search of the global minimum \cite{bo1,bo2}.  With this procedure, the minimization of $\cal F$ for each value of $n$ requires $15 \times {\cal N} = 45$ quantum dynamics calculations for H$_3$ and $15 \times {\cal N} = 90$ calculations for OH$_3$. 

\begin{figure}[!ht]
\includegraphics[width=\columnwidth]{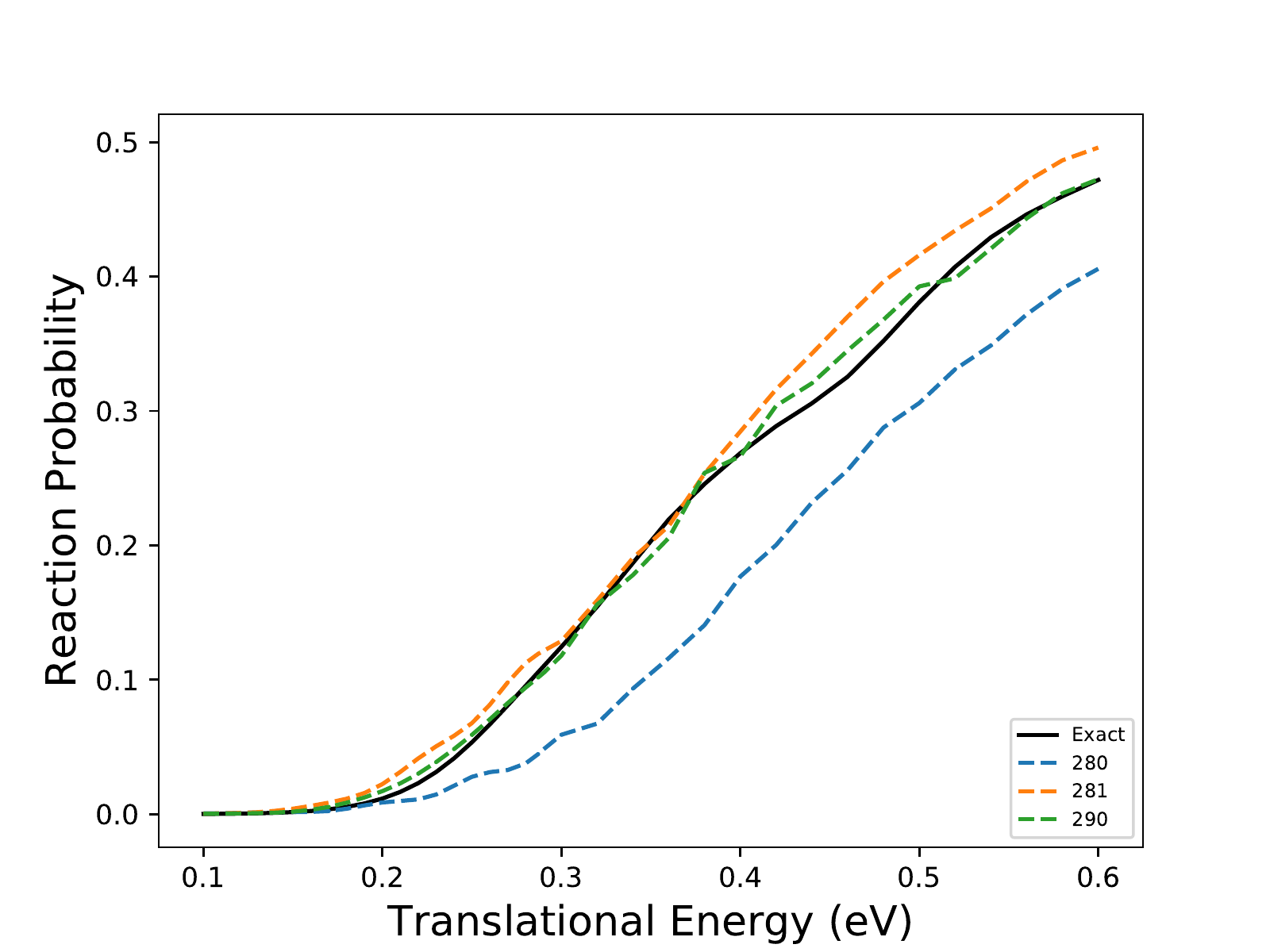} 
\caption{\label{fig2} 
 The reaction probability for the OH + H$_2$ $\rightarrow$ H + H$_2$O reaction as a function of the collision energy. The black solid curve -- accurate calculations from Ref. \cite{OH3-pes} based on the surface constructed with $\sim$17000 {\it ab initio} points. The dashed curves -- calculations based on the GP PES obtained with 280 {\it ab initio} points (blue); 281 points (orange) and 290 points (green). The RMSE and maximum error of the 290-point result are 0.0076 and 0.0195, respectively.   
}
\end{figure}

Figure 1a illustrates the performance of this algorithm in search of the best PES for reaction (5). As can be seen, the starting model $\cal G$ of the PES based on 22 randomly generated {\it ab initio} points produces highly inaccurate results, but the BO scheme converges to the correct PES after only 8 iterations ($8 \times 15 \times 3 = 360$ scattering calculations), yielding accurate results for the reaction probabilities (green dashed line) with the GP model  $\cal G$ of the PES  trained by 30 {\it ab initio} points. Figure 1b shows the model $\cal G$ of the PES obtained with $n = 30$ {\it ab initio} points, illustrating that Eqs. (\ref{sum-representation}) and (\ref{weights}) produce a physical surface.   

Figure 2 illustrates the performance of this algorithm for the 6D reaction (6). As the dimensionality of the configuration space increases, so does $n$ in Eq. (\ref{sum-representation}) required to represent acurately the PES. Nevertheless, accurate results for the reaction probabilities (green dashed line) are obtained with $n=290$ {\it ab initio} points, much smaller than the set of $\sim 17,000$ points used in previous work \cite{OH3-pes} to construct the PES with a NN fit. 
This result is obtained after 10 iterations (600 scattering calculations). 
 The RMS error of the reaction probabilities thus obtained is 0.0076. 
Note that, as any supervised learning technique, this algorithm is guaranteed to become more accurate when trained by more {\it ab initio} points. 

The efficiency of BO is perhaps best illustrated by the dependence of the RMSEs of the reaction probabilities as functions of the number of BO iterations, presented in Figure 3. 
Both panels of Figure 3 show that BO quickly decreases the RMSE of the reaction probabilities. In the upper panel showing the results for reaction (5), the initial surface is constructed 
using 22 {\it ab initio} points chosen at random and BO starts with $n = 23$. 
In the lower panel showing the results for reaction (6), the initial surface is constructed using 280 {\it ab initio} points chosen at random in the 6D space and BO (results represented by circles) starts with $n = 281$. To illustrate quantitatively the accuracy gain enabled by BO, we also show in the lower panel of Figure 3 the RMSE of the reaction probabilities computed with the PES obtained based on the corresponding number of points chosen at random and distributed in the configuration space using the Latin Hypercube Sampling (LHS) scheme to avoid clustering. 
The difference in the computation difficulty between the results shown by squares and circles in the lower panel of Figure 3 is 14 scattering calculations, i.e. the results shown by squares require one scattering calculations, whereas the results shown by circles require 15 scattering calculations. The key result illustrated by Figure 3 is that the convergence of BO is fast and monotonous, reducing the error of the scattering calculations dramatically after only two to three iterations for both reactive systems.  
This underscores the critical importance of BO for the approach proposed here and also underlines the huge potential of BO for other applications involving the inverse scattering problem. 

\begin{figure}[!ht]
\includegraphics[width=\columnwidth]{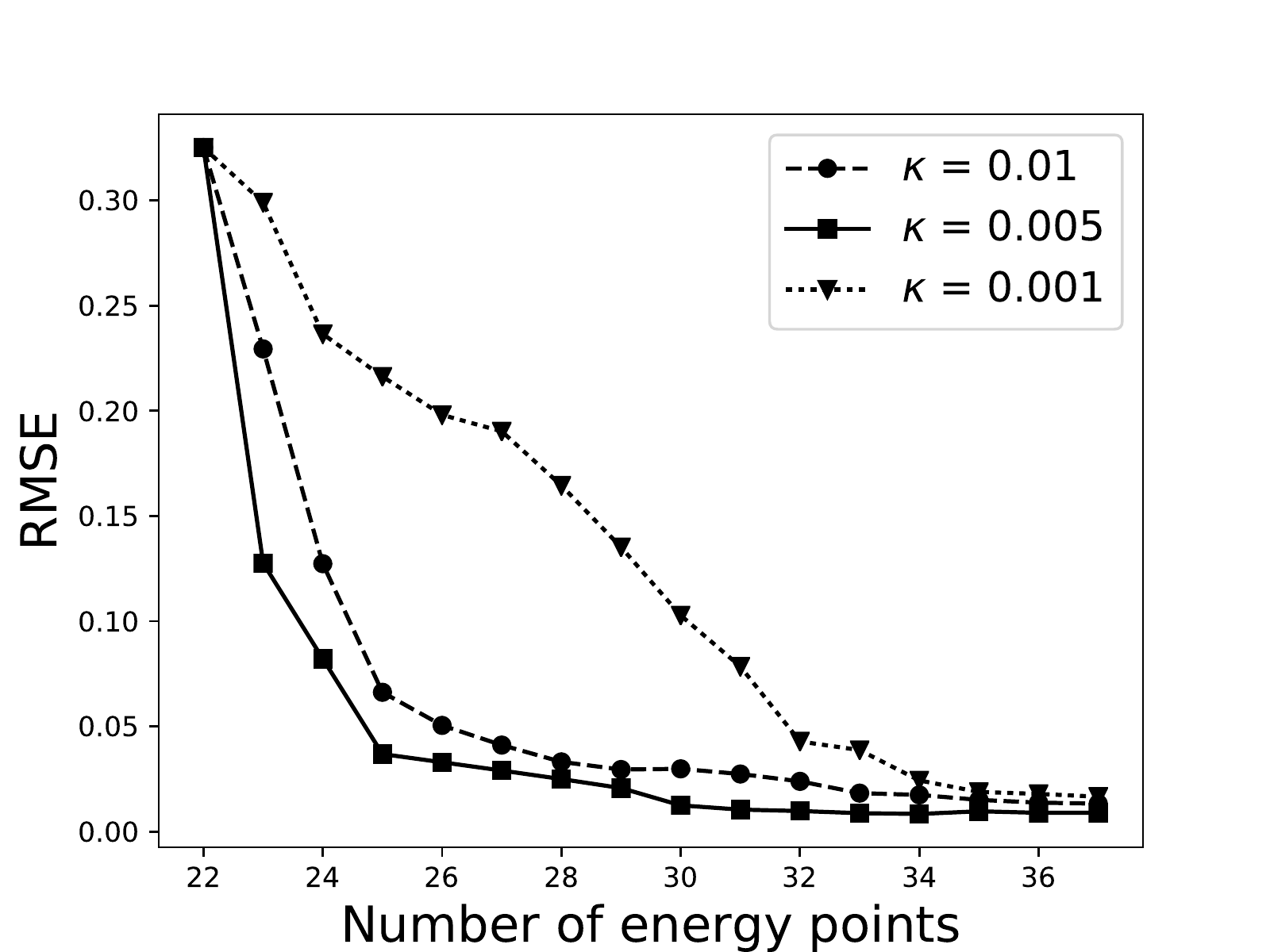} 
\includegraphics[width=\columnwidth]{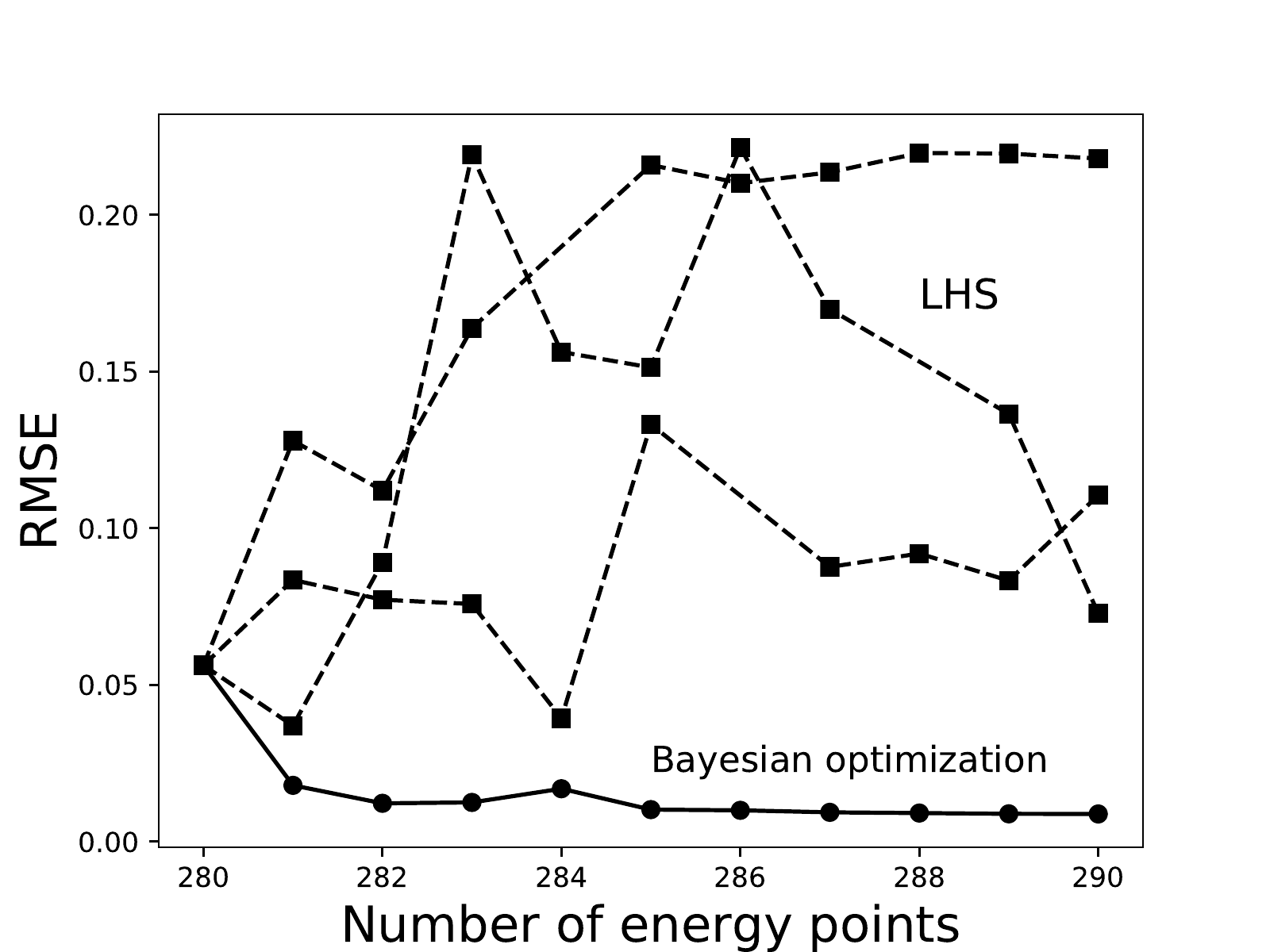} 
\caption{\label{fig2} 
Convergence of the RMSE of the reaction probabilities. Upper panel: the H$_2$ + H $\rightarrow$ H + H$_2$ reaction. 
Bayesian optimization starts at 22 points. 
The results shown by circles ($\kappa = 0.01$), squares ($\kappa = 0.005$) and triangles ($\kappa = 0.001$) illustrate the effect of the value of $\kappa$ on convergence of Bayesian optimization. Lower panel: the OH + H$_2$ $\rightarrow$ H + H$_2$O reaction. 
 Bayesian optimization of the results shown by circles starts at 280 points. 
The squares show the results of the scattering calculations with the PES obtained based on the corresponding number of points chosen at random and distributed in the configuration space using the Latin Hypercube Sampling scheme to avoid clustering. 
}
\end{figure}

\section{Obtaining PES without known dynamical results} 
 
If the quantum dynamics results are not known, the present method can be applied to construct accurate PES with a small number of {\it ab initio} energy points. Since GP regression is a supervised learning algorithm, GP models of PES become necessarily more accurate when trained by more {\it ab initio} points (i.e. any interpolation method becomes infinitely accurate in the limit of infinite number of points to be interpolated). The method proposed here aims to reach this limit with as few steps as possible. To achieve this, it is necessary to restrict the configuration space of the molecule to a reasonable volume and maximize the difference between of the results of subsequent iterations in Eq. (2). 

To illustrate the validity of this assumption, we show in Figure 4 a series of computations as functions of $n$, showing the convergence of the iterative calculations to the accurate results (black solid curve). Figure 4 shows that the optimization loop converges to the accurate PES after 48 iterations. 
We emphasize that the accurate results (black curve) were not used in any way in this calculation.

\begin{figure}[!ht]
\includegraphics[width=\columnwidth]{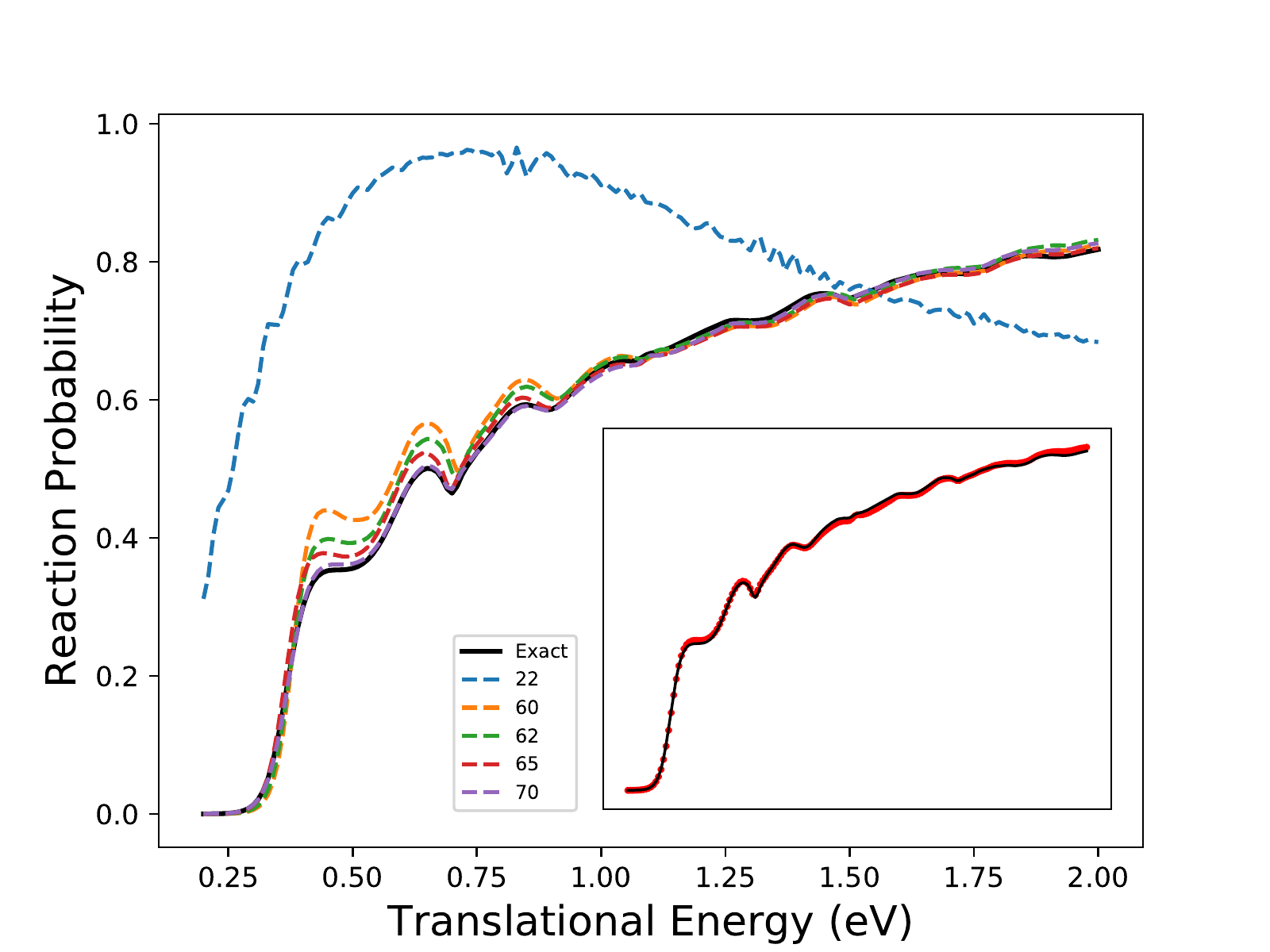} 
\caption{\label{fig3}
The reaction probabilities for the H$_2$ + H $\rightarrow$ H + H$_2$ reaction as functions of the collision energy. The black solid curve -- accurate calculations from Ref. \cite{H3-pes}.  The dashed curves -- the results of iterative calculations maximizing the difference between the reaction probabilities in successive iterations. The black curve is {\bf not used} for these calculations. 
The inset shows the agreement between the reaction probabilities (red symbols) based on the GP approach after 48 iterations (total of 70 {\it ab initio} points) and the exact results. 
}
\end{figure}

\section{Reproducing arbitrary scattering observables}

 Here, we extend the previous sections to construct a PES that, when used in quantum scattering calculations, reproduces an arbitrary set of observables.   We first modify the exact scattering results of Figure 1 by shifting along the energy axis and randomly modulating the black curve. This produces an arbitrary energy dependence of the reaction probabilities shown by the dot-dashed curve in Figure 5. The goal is to construct a PES that reproduces these arbitrarily chosen reaction probabilities.  
Note that the dot-dashed curve extends the interval of energies, where the reaction probability is zero, which means that the PES for this reaction must have a higher reaction barrier and cannot be reproduced with the original PES for H$_3$. Since this approach builds the PES based on the observables and is designed to yield the PES reproducing the observables, it is equivalent to solving the inverse scattering problem.

%

  We assume that a small ensemble of $E_i$ is known from some (not necessarily accurate) quantum chemistry calculation. As before, this ensemble serves as a starting point for the model $\cal G$ of the PES. However, in order to allow for the improvement of the PES, we now allow the energy of each point of the PES to be a variable $\varepsilon(\bm r)$. This variable is a function of $\bm r$ and allows the model $\cal G$ to sample from an interval of energies centered about the points of the initial rough PES.  The results of the scattering calculations are now used to train the model ${\cal F} [{\cal G}(n+1,  \varepsilon(\bm r), {\bm r} )]$, effectively increasing dimensionality of the variable space to ${\cal N} + 1$. 
 Figure 5 shows that this algorithm converges to the arbitrarily modified reaction probabilities after 32 iterations, producing a PES depicted in the lower panel. 

\begin{figure}[!ht]
\includegraphics[width=\columnwidth]{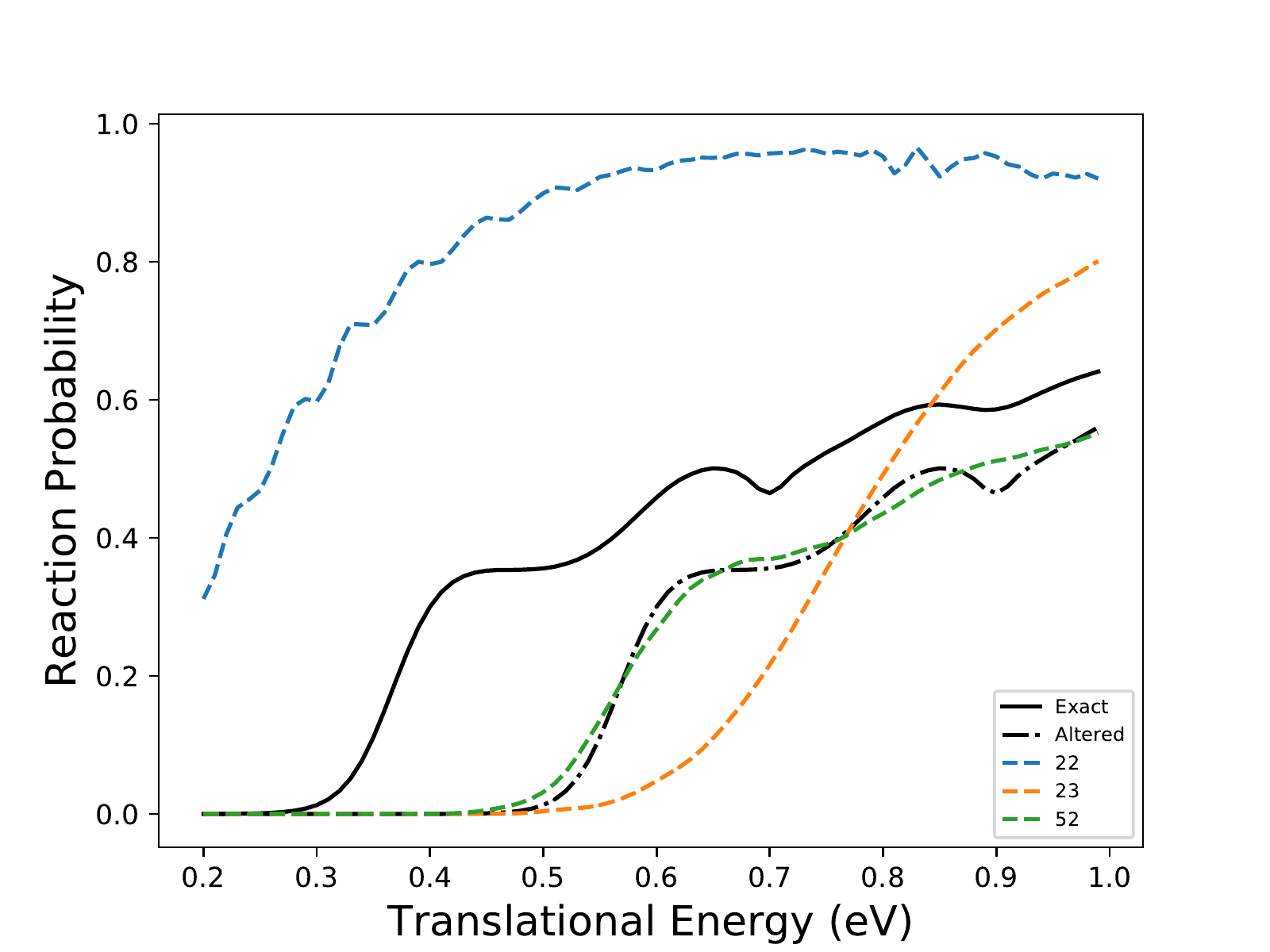} \\
\includegraphics[width=1\columnwidth]{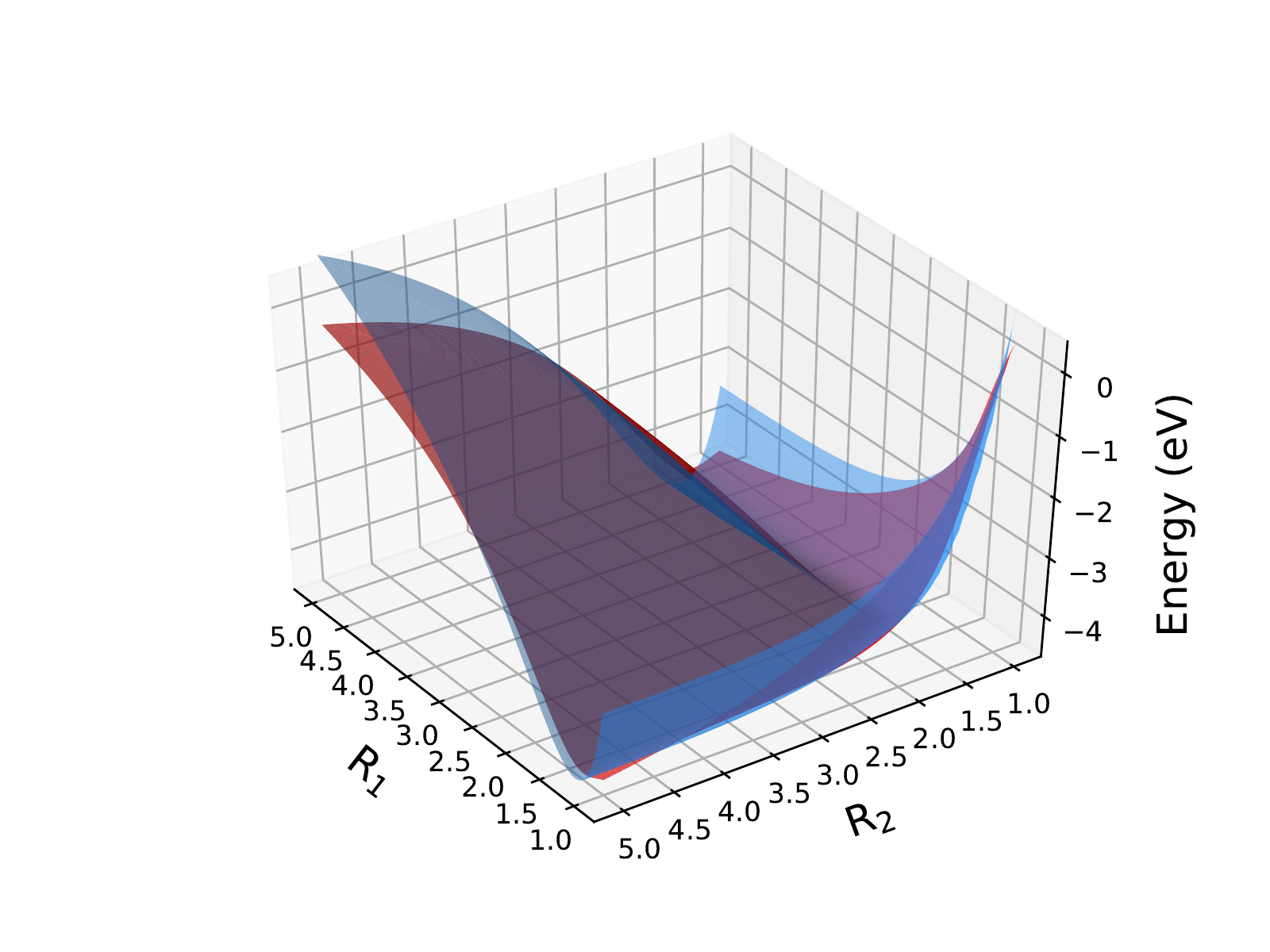} 
\caption{\label{fig4} 
Upper panel: The reaction probabilities for the modified H$_2$ + H $\rightarrow$ H + H$_2$ reaction as functions of the collision energy. The black dot-dashed curve is obtained  by a modification of the previous results (black solid curve) involving a translation along the energy axis. The ML models are trained to obtain the PES that would describe the new  reaction probabilities. The green dashed curve is a results of such training after 30 iterations, resulting in a surface constructed with 52 {\it ab initio} points. 
Lower panel: Comparison of the original PES (blue)  with the new PES (red) found by the BO algorithm. The new PES yields the reaction probabilities described by the green dashed curve in the upper panel. The RMSE and maximum error of the results shown by the green dashed curve are 0.016 and 0.0674, respecively. $R_1$ and $R_2$ are defined as in Figure 1.  
}
\end{figure}


\section{Conclusion}  

This work shows that it is possible to implement the feedback loop \\
\begin{minipage}{0.9\columnwidth}
\hspace{5.cm}
\begin{tikzpicture}[->,scale=.7]
   \node (j) at (-30:1cm) {\small PES };
   \node (k) at (210:1cm) {\small \hspace{-4.cm} Quantum Reaction Probabilities};
   \draw (-50:1cm) arc (-50:-130:1cm);
   \begin{scope}[xshift=0.1cm, yshift=0.3cm]
  \draw (210:1cm) arc (180:0:0.8cm);
   \end{scope}
\end{tikzpicture}
\end{minipage} \\
to build the global PES for  3D and 6D reaction systems.  The unique feature of this method is that the PES is built from a small number of {\it ab initio} points. 
At each iteration, an {\it ab initio} point is added to the most relevant part of the configuration space
and the global PES is automatically morphed to become more accurate. By construction, this approach produces PES yielding an accurate description of known reaction observables and the PES offers unique information on which parts of the interaction are most important for the outcome of the reaction process.

By construction, this method illustrates the lowest number of potential energy points (i.e. the minimum information) required for the non-parametric cosntruction of global PES for quantum reactive scattering calculations.  We showed that accurate quantum reactive scattering results can be obtained with 30 {\it ab initio} points for the 3D H + H$_2$ $\rightarrow$ H$_2$ + H reaction and 290 points for the 6D OH + H$_2$ $\rightarrow$ H + H$_2$O. For practical applications, the locations of these few points can be determined as described above with a low-level {\it ab initio} method and an approximate dynamical approach.  Once the positions of the points are known, this small number of {\it ab initio} points can be calculated with extremely high precision.



We emphasize that the method can also be used to construct accurate PES even if observables are not known. We showed that an algorithm based on {\it maximizing the difference} of computed observables in successive iterations  quickly converges to an accurate PES without the {\it a priori} knowledge of the full surface or dynamical results. The convergence to the correct limit is guaranteed by the observation that the ML model adopted here produces the correct surface in the limit of a large number of {\it ab initio} points.




The efficiency of the approach proposed here is limited by the difficulty of the dynamical calculations. However, one can use approximate dynamical methods for step (iii) in Eq. (2). This will result in PES, which yields an accurate description of the observables if used with this specific dynamical method. The error of this PES will be designed to compensate for the error of the dynamical calculations. While it remains to be seen if such an approach has predictive power outside the range of the observables used to construct the PES, we believe, our work opens the possibility of applying approximate dynamical methods such as the coupled states approach \cite{cs,cs-1,cs-2,cs-3,cs-4} for quantitative predictions of molecular scattering observables. 

Finally, we note that the approach presented here is general and can be used to construct the microscopic interaction potentials for any process with a 
$$
{\rm Microscopic~Interaction~Potential~~\Rightarrow~~Observable}
$$
dependence. Within the framework of the method proposed here, one should represent the interaction potential by a GP model $\cal G$ and the observable by another GP model ${\cal F}[{\cal G}]$. The GP ${\cal F}[{\cal G}]$ can then be used as the emulator function for Bayesian optimization to find the optimal model $\cal G$, leading to the most accurate description of the observable.  We believe this strategy can be applied to a wide range of problems in molecular physics.


\section{Acknowledgments}

This work is supported by NSERC of Canada. R.~V.~Krems also acknowledges support from the Chinese Academy of Sciences in the form of a visiting fellowship.
D. H. Zhang acknowledges support from the National Natural Science Foundation of China (grants 21433009, 21590804, 21688102), and the Strategic Priority Research Program of the Chinese Academy of Sciences (Grant No. XDB17000000).


\bibliography{reference}

\end{document}